\documentclass[final]{elsart}

\usepackage[dvips]{graphicx}

\usepackage{amssymb}
\usepackage{amsmath}

\begin{document}

\begin{frontmatter}

\title{Extreme high energy proton-gamma discrimination from space observations}

\author[UNAM]{A.D. Supanitsky\corauthref{cor}} and
\author[UNAM]{G. Medina-Tanco}
\address[UNAM]{Instituto de Ciencias Nucleares, UNAM, Circuito Exteriror S/N, Ciudad Universitaria,
M\'exico D. F. 04510, M\'exico.}
\corauth[cor]{Corresponding author. E-mail: supanitsky@nucleares.unam.mx.}

\begin{abstract}
The origin of the highest energy cosmic rays is still unknown. At present, the major uncertainties are located at
energies above $\sim 10^{19.5}$ eV, the expected beginning of the GZK suppression. This is mainly due to the low
statistics available, a problem that will be addressed in unprecedented way by the upcoming orbital detectors like
JEM-EUSO. The detection of very high energy photons is of great relevance for the understanding of the origin of
this extreme energy cosmic rays (EECR), due to the astrophysical information content. However, their discrimination
is an experimental challenge for current and future cosmic ray detectors. In this work we study the statistical
separation between hadron and photon showers from space observations at energies where both, the Landau-Pomeranchuk-Migdal
(LPM) effect and magnetospheric interactions are important for the development of the cascades. We base our analysis
on the $X_{max}$ parameter, which is already a well known composition discrimination parameter for ground based
fluorescence observatories. Our analysis applies to orbiting detectors in general. Nevertheless, we exemplify the
practical utilization of our technique by estimating a general upper limit to the photon fraction in the integral flux,
attainable by an ideal orbital detector with characteristics similar to JEM-EUSO. In the process we describe the resultant
asymmetry in the photon-hadron discrimination efficiency in galactic coordinates.
\end{abstract}

\end{frontmatter}

\section{Introduction}
\label{Int}

The cosmic ray flux must present, at least, a minor component of ultra high energy photons which may receive contributions 
from different sources. Besides the expected flux generated by the propagation of the EECR in the intergalactic medium 
\cite{Gelmini:07}, photons may also be originated in different astrophysical environments, as by-products of particle 
acceleration in nearby cosmic ray sources (see e.g. Ref. \cite{CentaurusA:08}) and, fundamentally, in top-down scenarios, 
even if not currently favored, involving the decay of super heavy relic particles or topological defects \cite{Aloisio:04}. 
Extreme energy photons have not been unambiguously observed yet. However, it is expected that planned space observatories, 
like JEM-EUSO \cite{Ebisuzaki:09,Gustavo:09,Yoshi:09} and S-EUSO \cite{SEUSO}, with their unprecedented exposure, change 
this situation in the next few years.

Ultra high energy photons can interact with the magnetic field of the Earth producing electron positron pairs which modify 
the development of photon initiated atmospheric showers (see Ref. \cite{Homola:07} and references there in). The probability 
of photon splitting depends on the intensity of the component of the magnetic field  perpendicular to the direction of 
propagation of the photon. Thus, the characteristics of the corresponding showers strongly depend on the geographical 
position of the impact point and the direction of incidence of the incoming primary photon. Additionally, the development 
of a photon shower is affected by the so-called LPM effect (Landau and Pomeranchuk \cite{Landau:53a,Landau:53b}, Migdal 
\cite{Migdal:56}), which consists in a reduction of the Bethe-Heitler cross section for pair production by photons, producing 
a delay in the development of the shower. Therefore, the longitudinal evolution of high energy photon showers is dominated 
by the interplay between these two effects which has a direct impact on the practical possibility of discrimination between 
photon and proton showers.   

In the present work we develop two complementary techniques in order to evaluate an upper limit on the
fraction of photons relative to proton primaries in the integral cosmic ray flux by using the atmospheric
depth of maximum development, $X_{max}$, of the corresponding atmospheric showers. As previously mentioned,
the characteristics of the photon showers strongly depend on the geographical coordinates of the impact point
of the shower; therefore, the translation of a space observatory along its orbit is a distinctive parameter
which adds richness and complexity to the analysis with respect to a traditional Earth bound observatory and
must be taken into account. We apply these new techniques to the case of an orbital detector similar to 
JEM-EUSO.

\section{Proton-gamma discrimination}

Given a sample of $N$ events, an ideal upper limit to the photon fraction may be calculated under 
the a priori assumption that actually no photon exists in the sample,
\begin{equation}
\mathcal{F}_{\gamma}^{min} = 1-(1-\alpha)^{1/N}
\label{Fuplideal}
\end{equation} 
where $\alpha$ is the confidence level of rejection. Eq. (\ref{Fuplideal}) has been already used to estimate the 
photon sensitivity in the experimental context \cite{Homola:09}. In practice, the probability of the existence of 
photons must be realistically assessed through some observational technique which involves the determination of 
experimental parameters which, in turn leads unavoidably to less restrictive upper limits than the previous one. 

In this work $X_{max}$ is used as a discrimination parameter between proton and photon showers. The program CONEX 
\cite{conex} is used to generate a library of proton and photon showers. The proton library consists of 
$\sim 4 \times 10^{5}$ showers following a power law energy spectrum of spectral index $\gamma = -2.7$ in the interval 
[$10^{19.3}, 10^{21}$] eV with uniformly distributed arrival directions. The photon library consists of more than 
$5.5 \times 10^{5}$ showers generated under the same previous conditions but the cores were now uniformly distributed 
on the surface of the Earth in order to properly take into account pre-showering (i.e., photon splitting) in the 
geomagnetic field.    

A quantitative assessment of the discrimination power of the $X_{max}$ parameter can be obtain through the merit factor,
\begin{equation}
\eta = \frac{\textrm{med}[X_{max}^\gamma]-\textrm{med}[X_{max}^{pr}]}{\sqrt{(\Delta X_{max}^\gamma)^2+(\Delta X_{max}^{pr})^2}}
\label{eatdef}
\end{equation}
where $\textrm{med}[X_{max}^A]$ is the median of $X_{max}^A$ ($A=\gamma, pr$) distribution and $\Delta X_{max}^A$ is one 
half of the length of the interval of 68\% of probability of $X_{max}^{A}$ distribution. 

\begin{figure}[!bt]
\centering
\includegraphics[width=12cm]{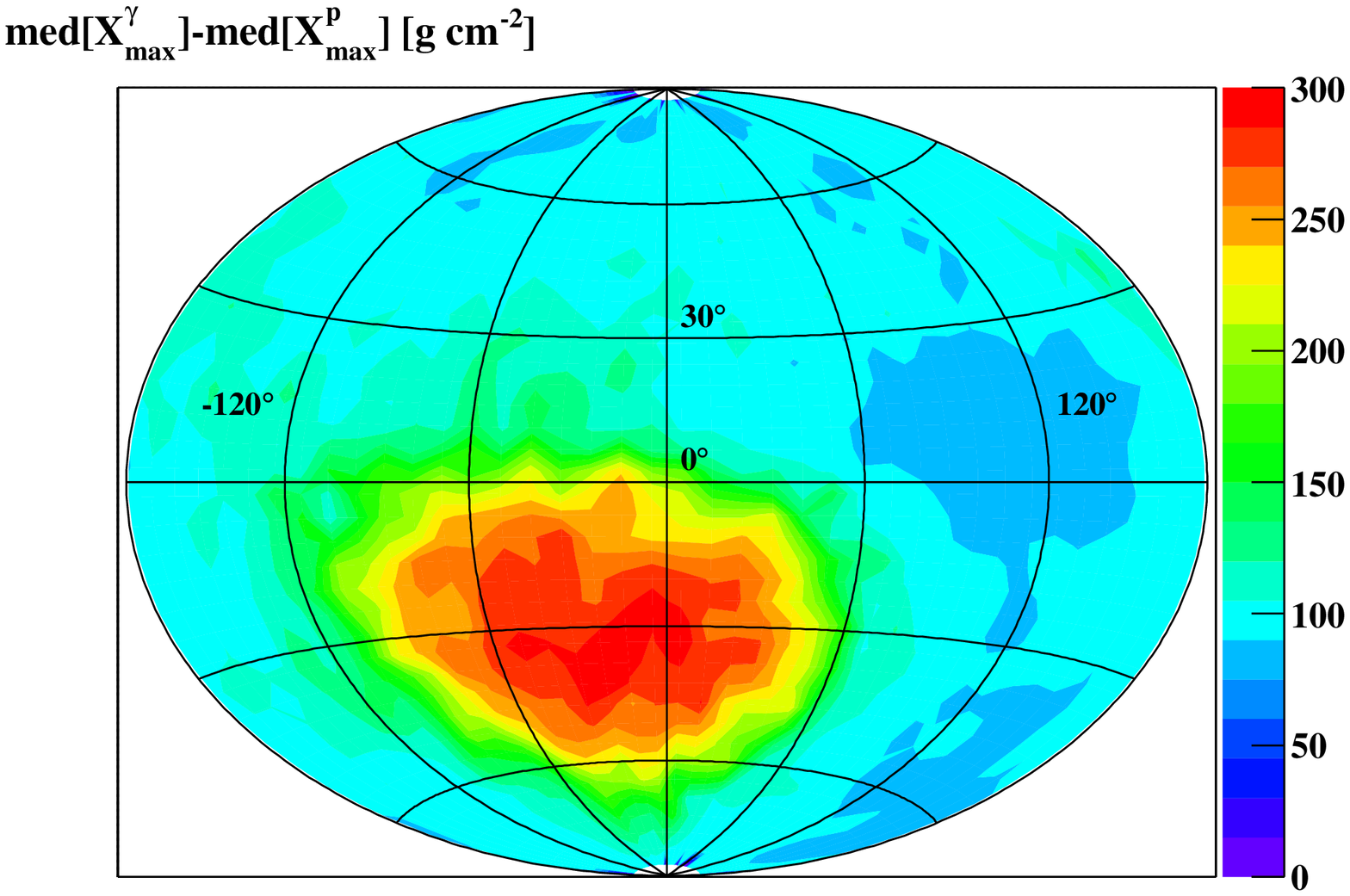}
\includegraphics[width=12cm]{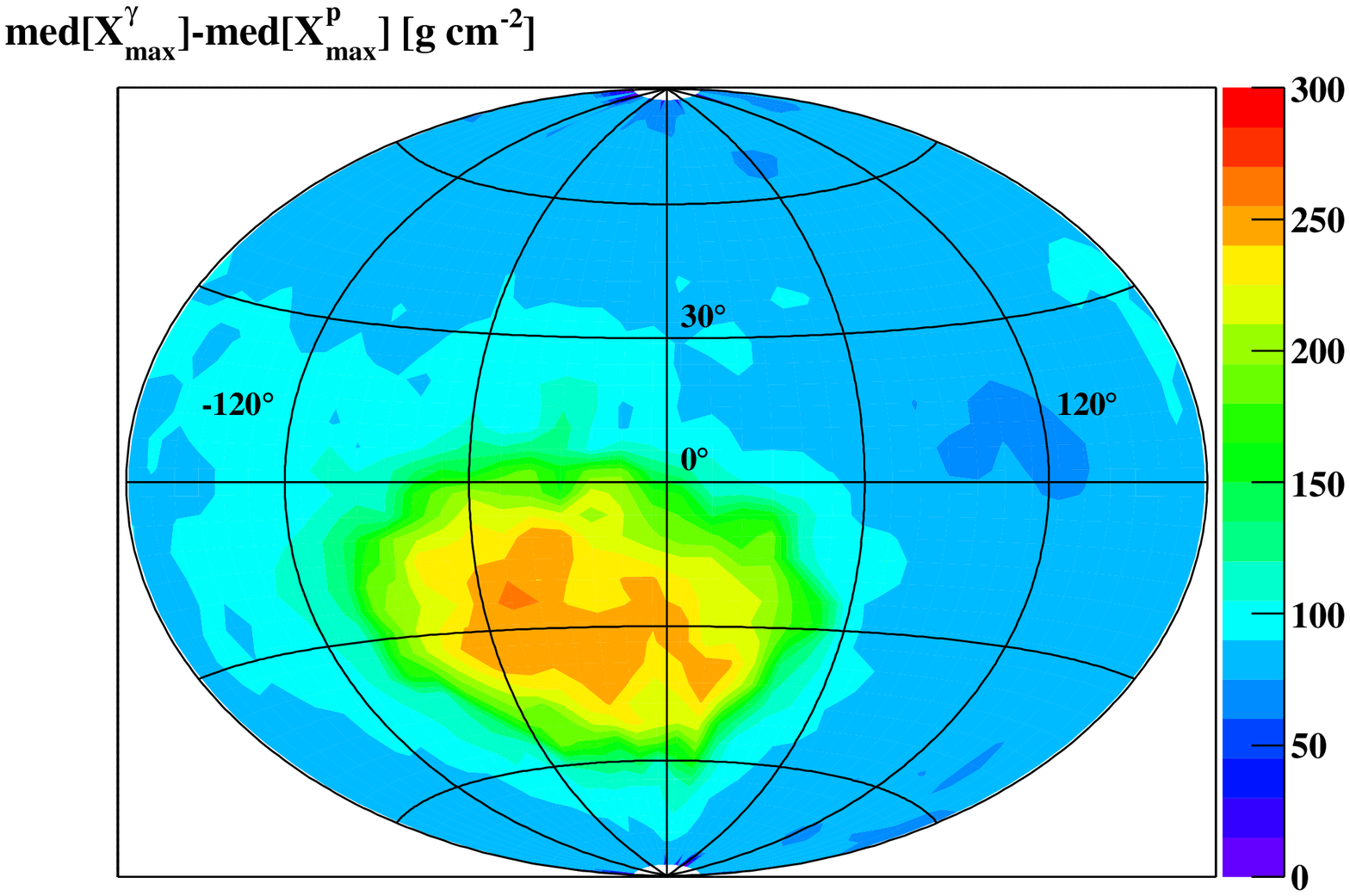}
\caption{Contour plots of $\textrm{med}[X_{max}^\gamma]-\textrm{med}[X_{max}^{pr}]$ as a function of latitude and 
longitude of the core location on the Earth surface. The showers considered are such that $\theta \in [30^\circ, 60^\circ]$ 
(top panel) and $\theta \in [45^\circ, 90^\circ]$ (bottom panel) and for $E \in [10^{19.8}, 10^{20}]$ eV.}
\label{Map}
\end{figure}

Fig. \ref{Map} shows, in Aitoff projection, a contour plot of $\textrm{med}[X_{max}^\gamma]-\textrm{med}[X_{max}^{pr}]$ 
as a function of latitude and longitude of the core on the Earth, for $\theta \in [30^\circ, 60^\circ]$ (top panel) and 
$\theta \in [45^\circ, 90^\circ]$ (bottom panel) and for $E \in [10^{19.8}, 10^{20}]$ eV. It can be seen that there are 
regions over the Earth surface where this difference is larger and in which the discrimination between protons and 
photons is more efficient. Note that for $\theta \in [30^\circ, 60^\circ]$, the difference between the medians of 
$X_{max}$ of the proton and gamma distributions is larger, this is due to the fact that, on average, photon splitting 
is more important for larger values of zenith angles. 

\begin{figure}[!bt]
\centering
\includegraphics[width=12cm]{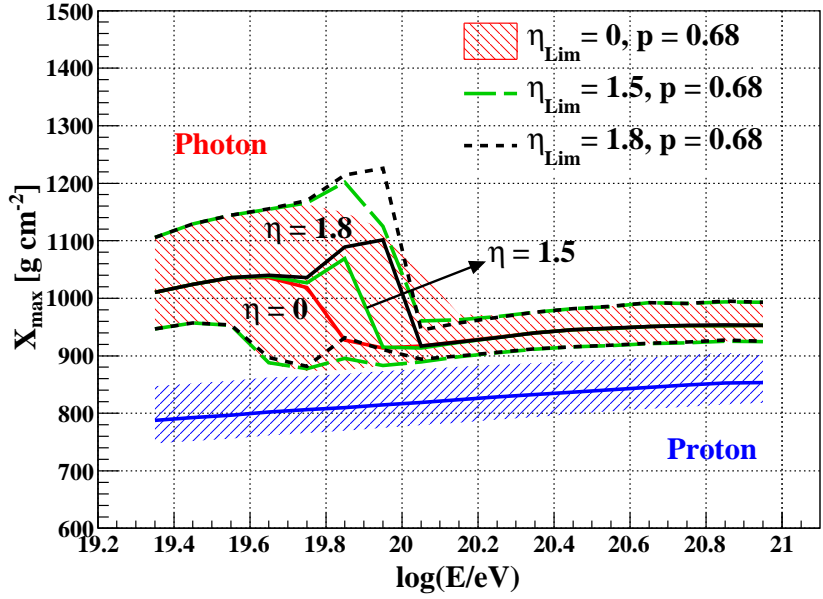}
\caption{Median and region of 68\% of probability of $X_{max}$ as a function of primary energy for 
$\theta \in [30^\circ, 60^\circ]$. In the case of photons different masks are considered, $\eta_{Lim}=0$ 
(all the events), $\eta_{Lim}=1$ and $\eta_{Lim}=1.5$.}
\label{XmaxCL}
\end{figure}

Motivated by this result the concept of mask, $\Omega(\eta_{Lim})$, is introduced as those regions over the Earth 
surface where $\eta$ is larger than a given value $\eta_{Lim}$. Fig. \ref{XmaxCL} shows the median of $X_{max}$ and 
the region of 68\% of probability as a function of primary energy for protons and photons with 
$\theta \in [30^\circ, 60^\circ]$. For the case of photons different masks are considered: $\Omega(\eta_{Lim}=0)$ 
(all events), $\Omega(\eta_{Lim}=1)$ and $\Omega(\eta_{Lim}=1.5)$. Note that the masks are functions of primary energy. 
From the figure it can be seen that, for photons, there are three well defined regions. The first corresponds to primary 
energies smaller than $\sim 10^{19.5}$ eV in which the $X_{max}^\gamma$ distribution is composed of LPM dominated showers. 
The second region extends from $\sim 10^{19.5}$ eV up to $\sim 10^{20.1}$ eV and is such that the $X_{max}^\gamma$ 
distribution is composed by LPM dominated showers that both, suffer and do not suffer photon splitting in the Earth 
magnetic field. In the third region all showers undergo photon splitting which generates, on average, smaller values 
of $X_{max}^\gamma$ and smaller fluctuations \cite{Vankov:03}.      

\begin{figure}[!h]
\centering
\includegraphics[width=11cm]{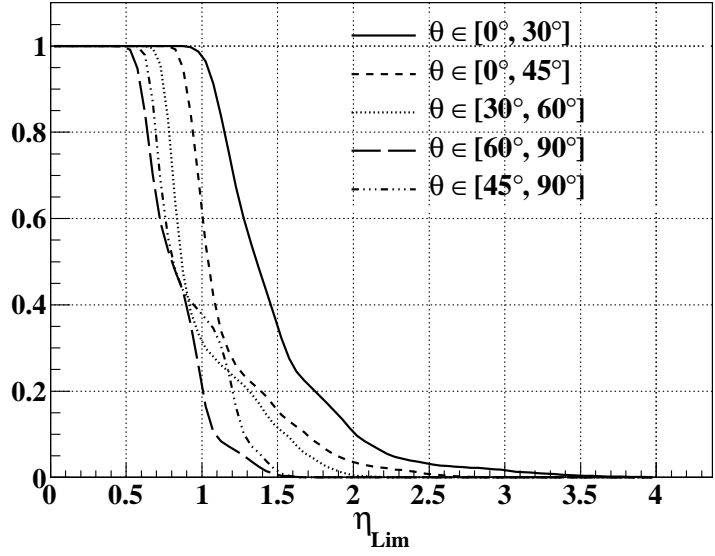}
\caption{Fraction of events as a function of $\eta_{Lim}$ for different cuts in zenith angle and for 
$E\in[10^{19.8},10^{20}]$ eV.}
\label{EtaLim}
\end{figure}

The $X_{max}^\gamma$ distributions obtained for masks with larger $\eta_{Lim}$ allow a better separation between 
protons and photons. However, the total number of events also depends on the assumed mask and, in particular, 
decreases with $\eta_{Lim}$. Fig. \ref{EtaLim} shows the fraction of events as a function of $\eta_{lim}$ for 
different cuts in zenith angle and for $E\in[10^{19.8},10^{20}]$ eV. It can be seen that for vertical showers 
larger values of $\eta$ are obtained which are also distributed over a wider range, again, the photon splitting 
is more important for larger values of zenith angle.     

Two methods are developed in order to calculate an upper limit for the photon fraction. The first one is based 
on the abundance estimator first introduced in Ref. \cite{Supanitsky:08},
\begin{equation}
\xi_{X_{max}} = \frac{1}{N} \sum_{i=1}^{N} P_{\gamma}(X_{max}^i)
\label{eatdef}
\end{equation}
where $X_{max}^{i}$ are experimental values of $X_{max}$, $N$ is the sample size and
\begin{equation}
P_{\gamma}(X_{max}) = \frac{f_\gamma(X_{max})}{f_\gamma(X_{max})+f_{pr}(X_{max})} 
\label{pgamma}
\end{equation}
with $f_\gamma(X_{max})$ and $f_{pr}(X_{max})$ the photon and proton distribution functions, respectively. 
$\xi_{X_{max}}$ is an estimator of the photon abundance, $c_{\gamma}=N_\gamma/N$ where $N_\gamma$ is the 
number of photons in the sample. The mean value of $\xi_{X_{max}}$ and the variance can be written as,
\begin{eqnarray}
\langle \xi_{X_{max}} \rangle (c_\gamma) &=& u_1 c_\gamma + u_2, \\
Var[\xi_{X_{max}}] &=& \frac{1}{N}\ (v_1 c_\gamma + v_2),
\end{eqnarray}
where $u_1=\alpha_1-\alpha_2$, $u_2=\alpha_2$, $v_1=\alpha_3-\alpha_4+\alpha_2^2-\alpha_1^2$ and 
$v_2=\alpha_4-\alpha_2^2$. Here 
\begin{eqnarray}
\alpha_1 &=& \int dX_{max}\ P_{\gamma}(X_{max})\ f_{\gamma}(X_{max}), \\
\alpha_2 &=& \int dX_{max}\ P_{\gamma}(X_{max})\ f_{pr}(X_{max}), \\
\alpha_3 &=& \int dX_{max}\ P_{\gamma}^2(X_{max})\ f_{\gamma}(X_{max}), \\
\alpha_4 &=& \int dX_{max}\ P_{\gamma}^2(X_{max})\ f_{pr}(X_{max}). 
\end{eqnarray}

For large enough values of $N$, $\xi_{X_{max}}$ is distributed as a Gaussian variable. Therefore, 
the region on the plane $\xi_{X_{max}}-c_\gamma$ of probability $p=\alpha$ is enclosed by the 
functions, 
\begin{equation}
\xi_\alpha^\pm (c_\gamma) = u_1\ c_\gamma+u_2 \pm \frac{s(\alpha)}{\sqrt{N}}\ \sqrt{v_1\ c_\gamma+v_2}, 
\end{equation}
$c_\gamma=0$ and $c_\gamma=1$. Here $s(\alpha)=\sqrt{2}\ \textrm{Erf}^{-1}(\alpha)$ where $\textrm{Erf}^{-1}(x)$ 
is the inverse of the error function, 
\begin{equation}
\textrm{Erf}(x) = \frac{2}{\sqrt{\pi}} \ \int^x_0 dt \ \exp\left( -\frac{t^2}{2}  \right).
\end{equation}

For the case in which $\xi_{X_{max}}$ is compatible with a pure proton sample, an upper limit to the photon fraction, 
$c_\gamma^{min}$, can be obtained solving the equation $\xi_\alpha^+ (c_\gamma=0) = \xi_\alpha^- (c_\gamma^{min})$, which 
gives,
\begin{equation}
c_\gamma^{min}=\frac{s(\alpha)}{u_1^2} \left( \frac{2\ u_1 \sqrt{v_2}}{\sqrt{N}} + \frac{s(\alpha)\ v_1}{N} \right).
\label{cgmin}
\end{equation}

The distribution functions needed to calculate $c_\gamma^{min}$ are obtained from the simulated data by using the 
non-parametric method of kernel superposition with adaptive bandwidth \cite{Silvermann:86,Supanitsky:08}.

\begin{figure}[!h]
\centering
\includegraphics[width=11cm]{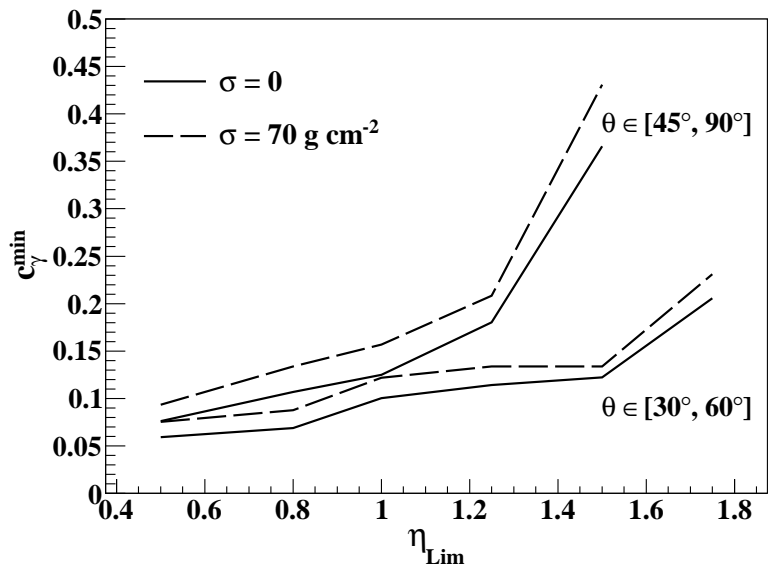}
\caption{$c_\gamma^{min}$ as a function of $\eta_{Lim}$ for $E\in[10^{19.8},10^{20}]$ eV, $\theta \in [30^\circ, 60^\circ]$ 
and $\theta \in [45^\circ, 90^\circ]$ and with and without assuming a Gaussian uncertainty of $70$ g cm$^{-2}$ on the 
reconstruction of $X_{max}$.}
\label{Cgmin}
\end{figure}

It can be seen from Eq. (\ref{cgmin}) that, the larger the sample size, the smaller $c_\gamma^{min}$. Although, it 
is not obvious from this expression, it is possible to show that for larger values of $\eta$ also smaller values of 
$c_\gamma^{min}$ are obtained. Fig. \ref{Cgmin} shows $c_\gamma^{min}$ for $\alpha=0.95$ as a function of $\eta_{Lim}$, 
i.e. for different masks, for $E\in[10^{19.8},10^{20}]$ eV, $\theta \in [30^\circ, 60^\circ]$ and 
$\theta \in [45^\circ, 90^\circ]$. In the figure, the solid lines show the results for the ideal case of no reconstruction 
errors, while the other curves were obtained under the assumption of a Gaussian uncertainty. In the latter case, errors 
with an uncertainty of $70$ g cm$^{-2}$ for $\theta \in [30^\circ, 60^\circ]$ and $\theta \in [45^\circ, 90^\circ]$ were 
assumed \cite{Berat:10}. It can be seen that $c_\gamma^{min}$ increases with $\eta_{Lim}$, which means that although the 
discrimination power of $X_{max}$ increases, the number of events decreases so rapidly producing larger values of 
$c_\gamma^{min}$, i.e. in this case the number of events is more important than the discrimination power of $X_{max}$ for 
a given mask. 

The second method developed here consists in finding a cut on $X_{max}$ which minimizes the expression of the upper 
limit obtained assuming a pure proton composition, 
\begin{equation}
\mathcal{F}_{UL}(X_{max}^c) = \frac{N_\alpha(N F_{pr}(X_{max}^c))}{N F_{\gamma}(X_{max}^c)},
\label{Fuplimcut}
\end{equation}
where $N_\alpha(n)$ is the upper limit on the number of photons $n$ at a confidence level $\alpha$ obtained assuming 
a Poisson distribution and,  
\begin{equation}
F_A(X_{max}^c)=\int_{X_{max}^c}^\infty d X_{max}\ f_A(X_{max}).
\end{equation}

\begin{figure}[!h]
\centering
\includegraphics[width=4.5cm]{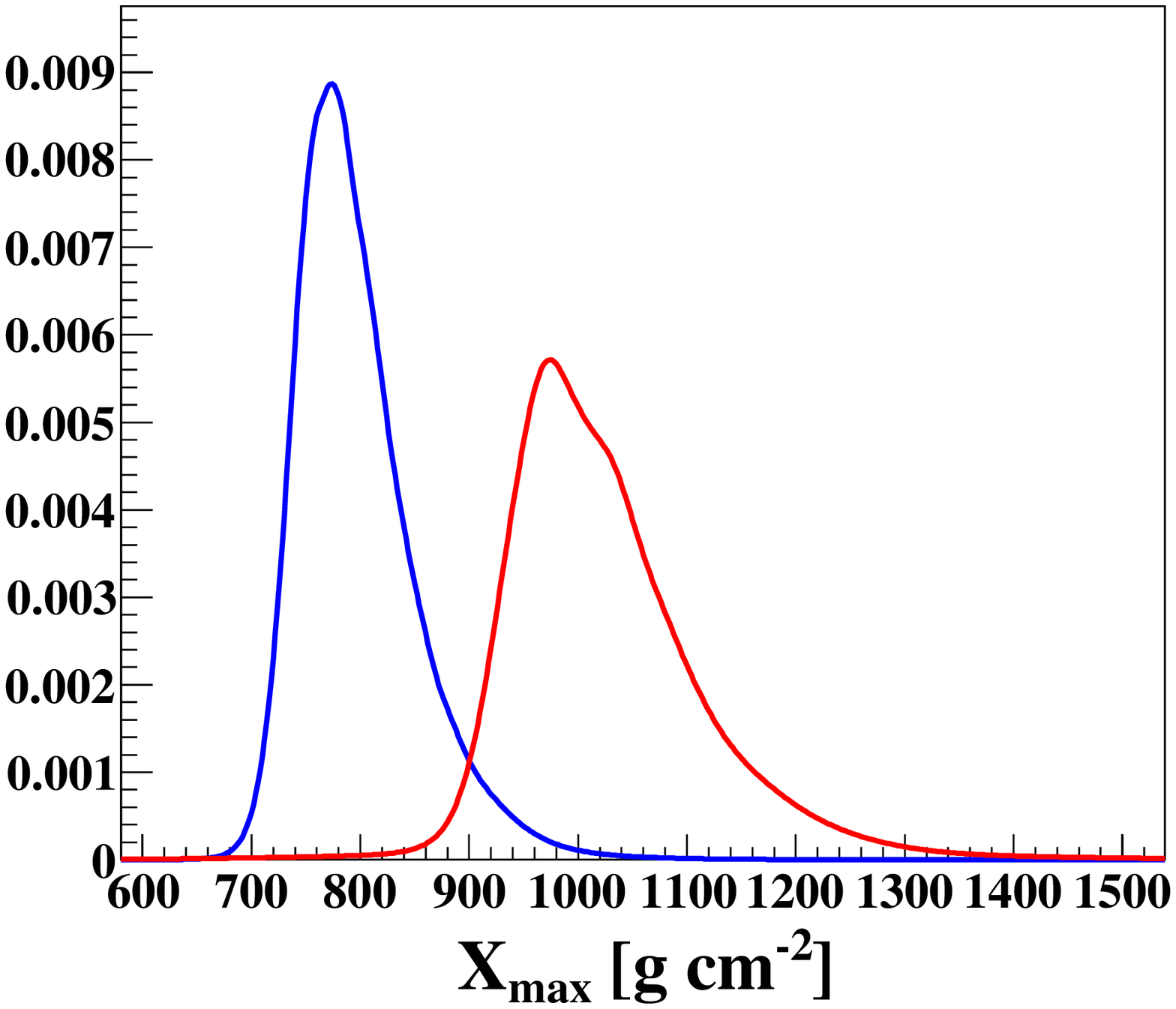}
\includegraphics[width=4.5cm]{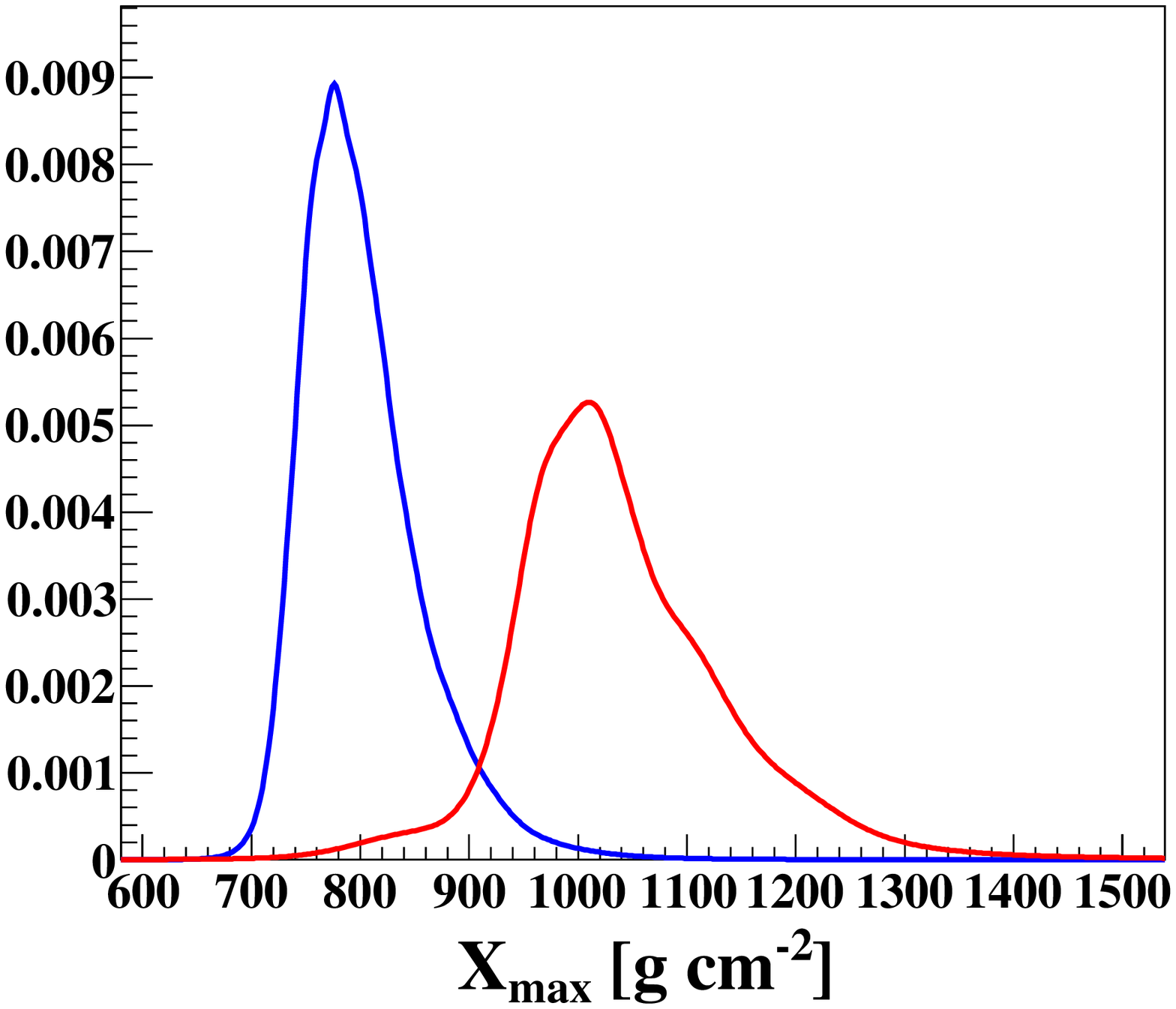}
\includegraphics[width=4.5cm]{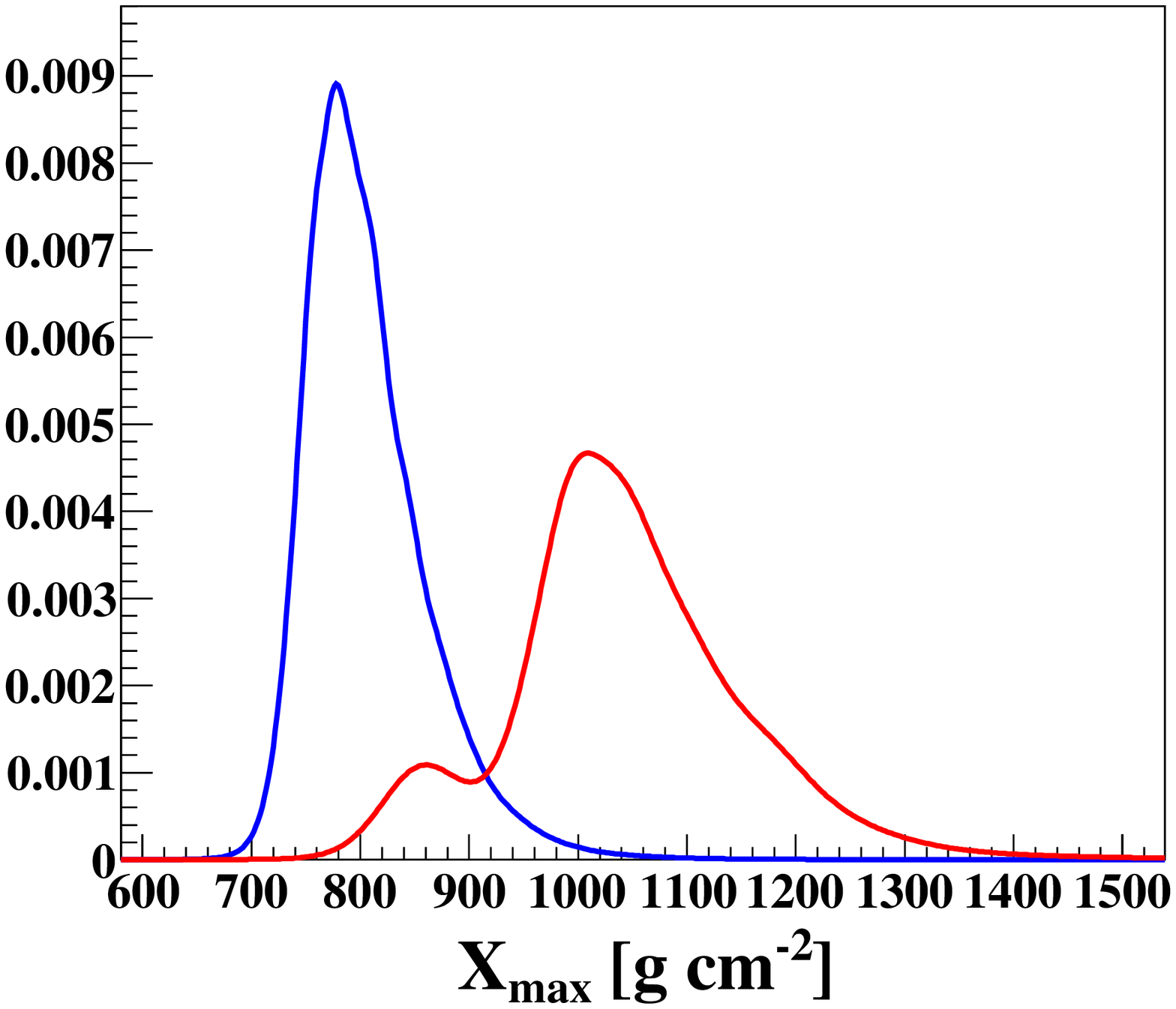}
\includegraphics[width=4.5cm]{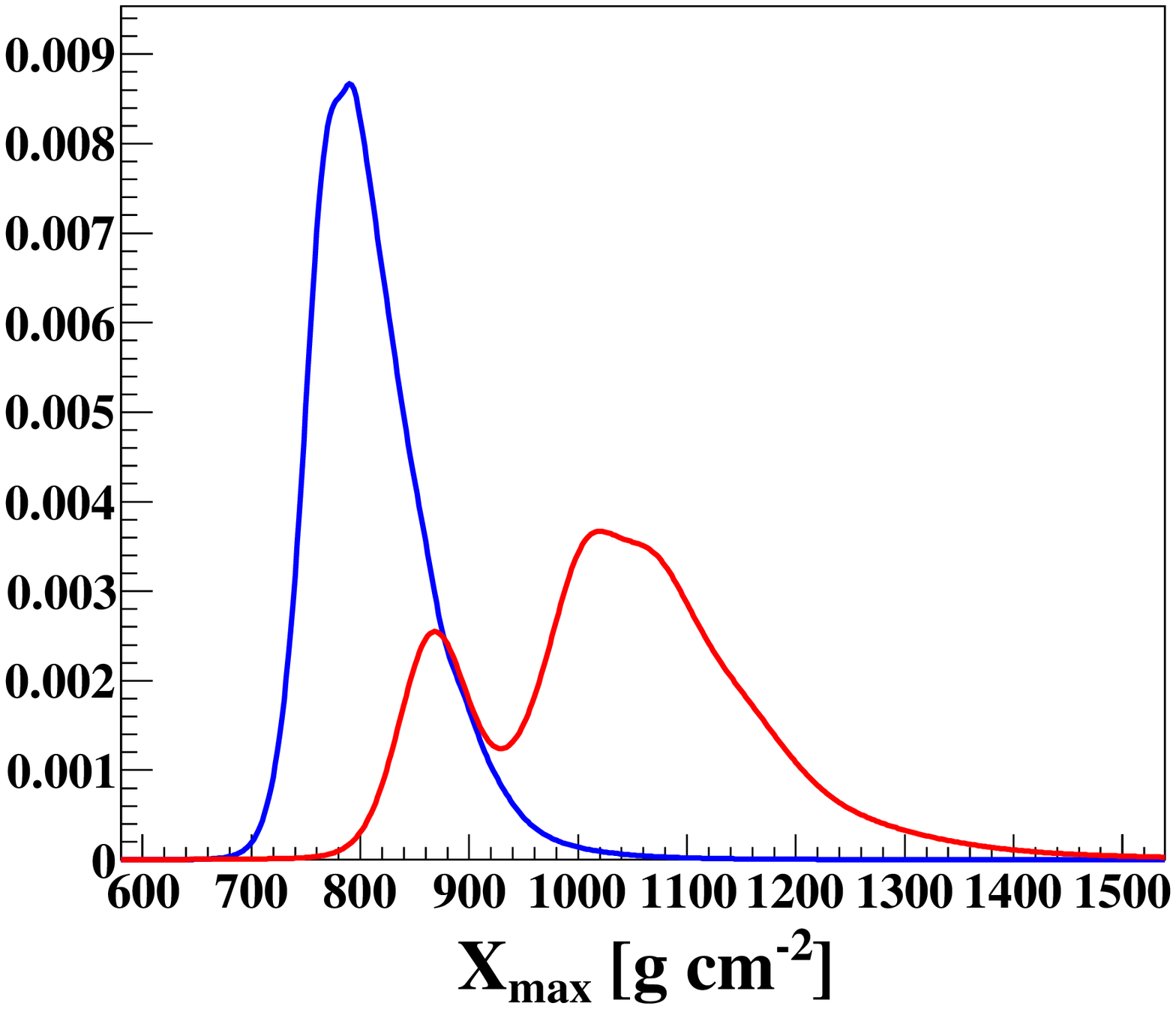}
\includegraphics[width=4.5cm]{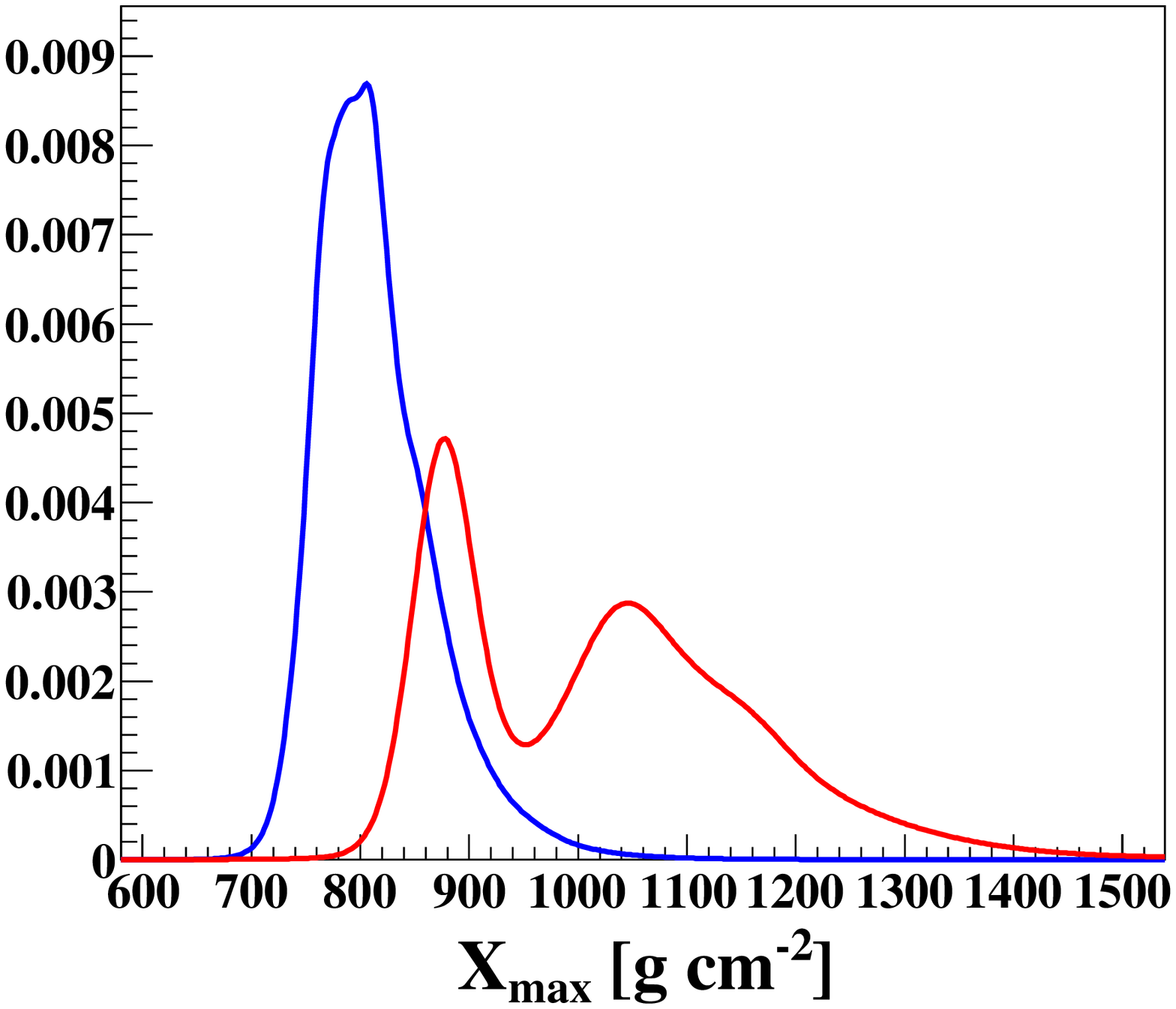}
\includegraphics[width=4.5cm]{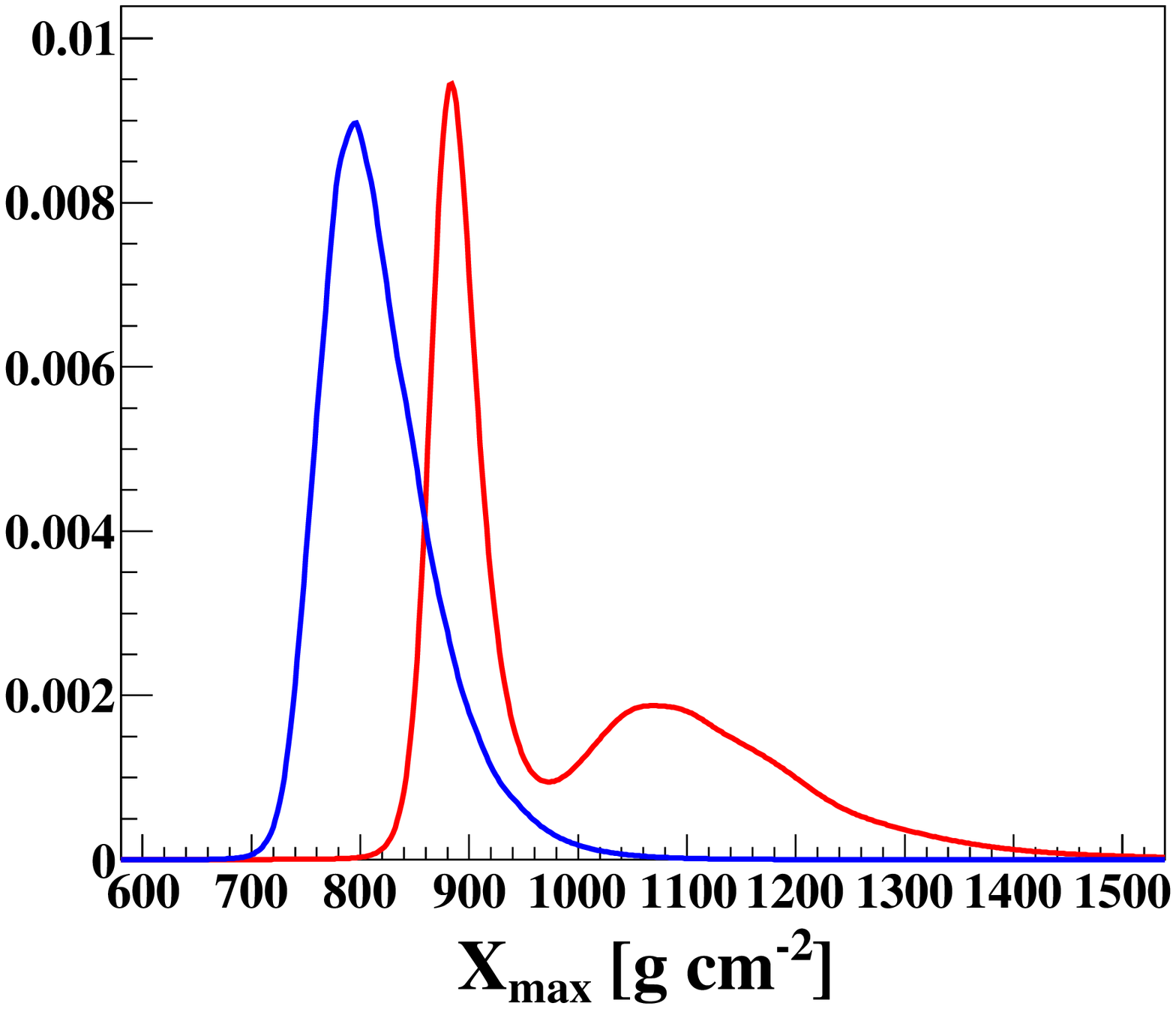}
\includegraphics[width=4.5cm]{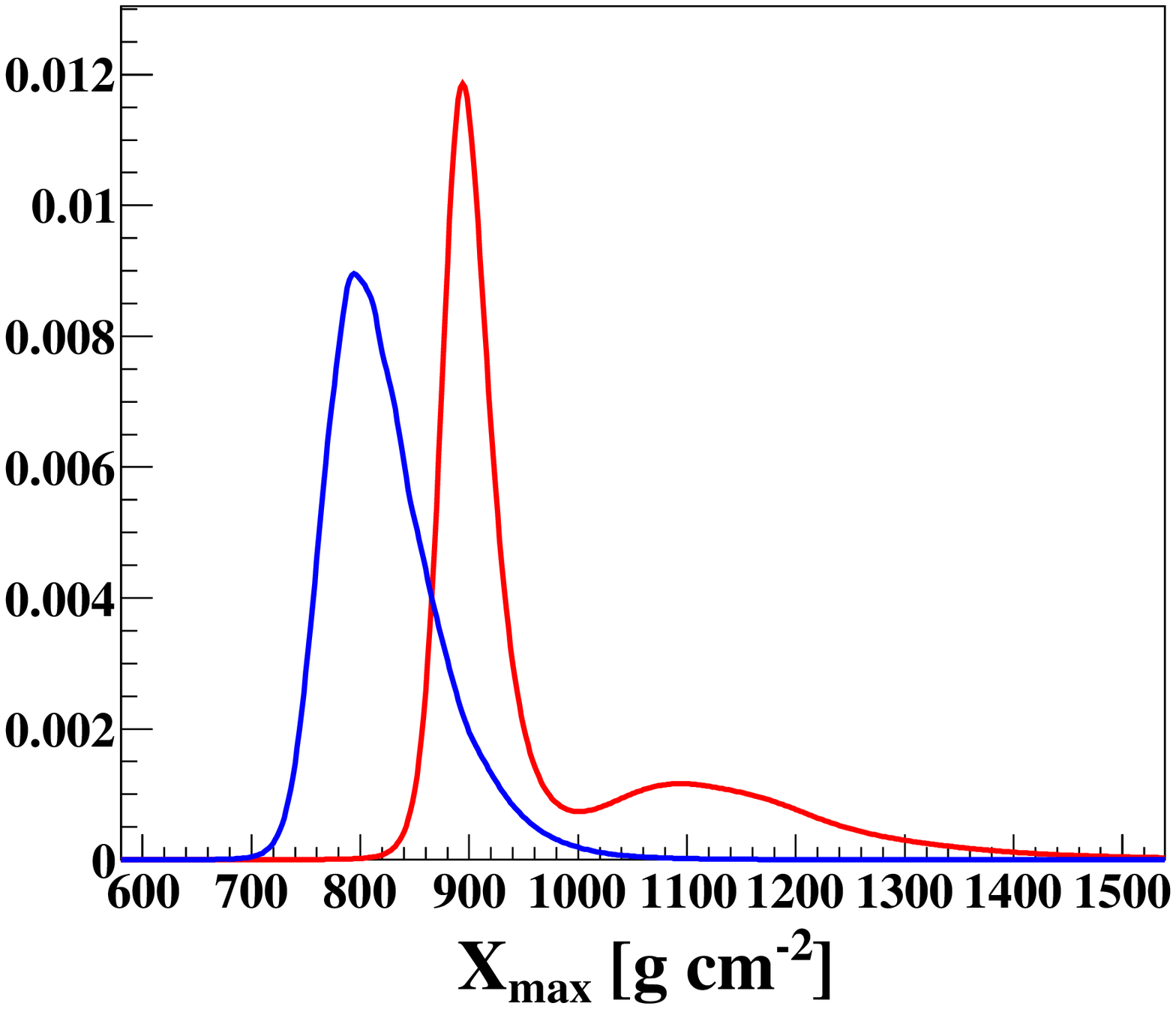}
\includegraphics[width=4.5cm]{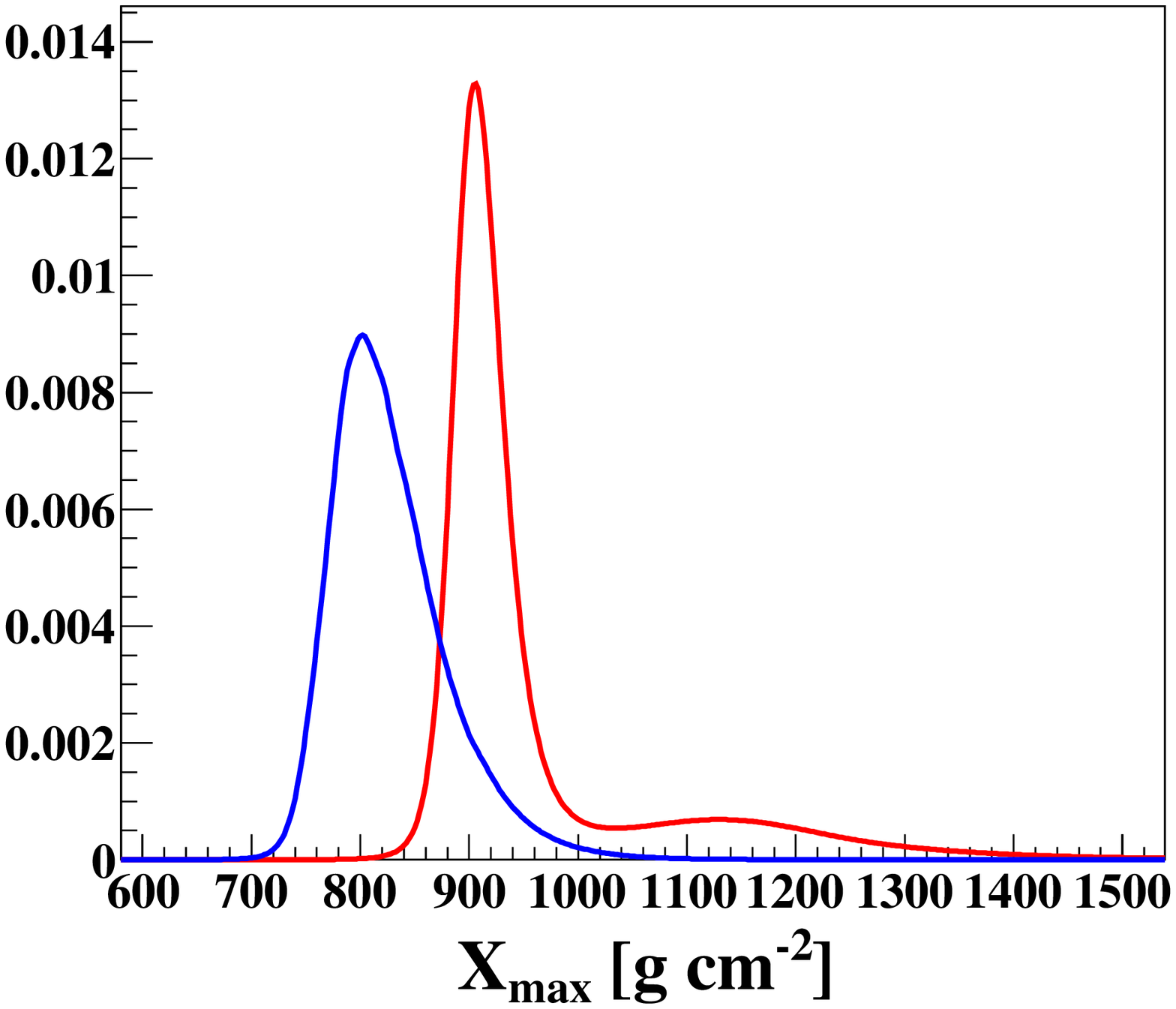}
\includegraphics[width=4.5cm]{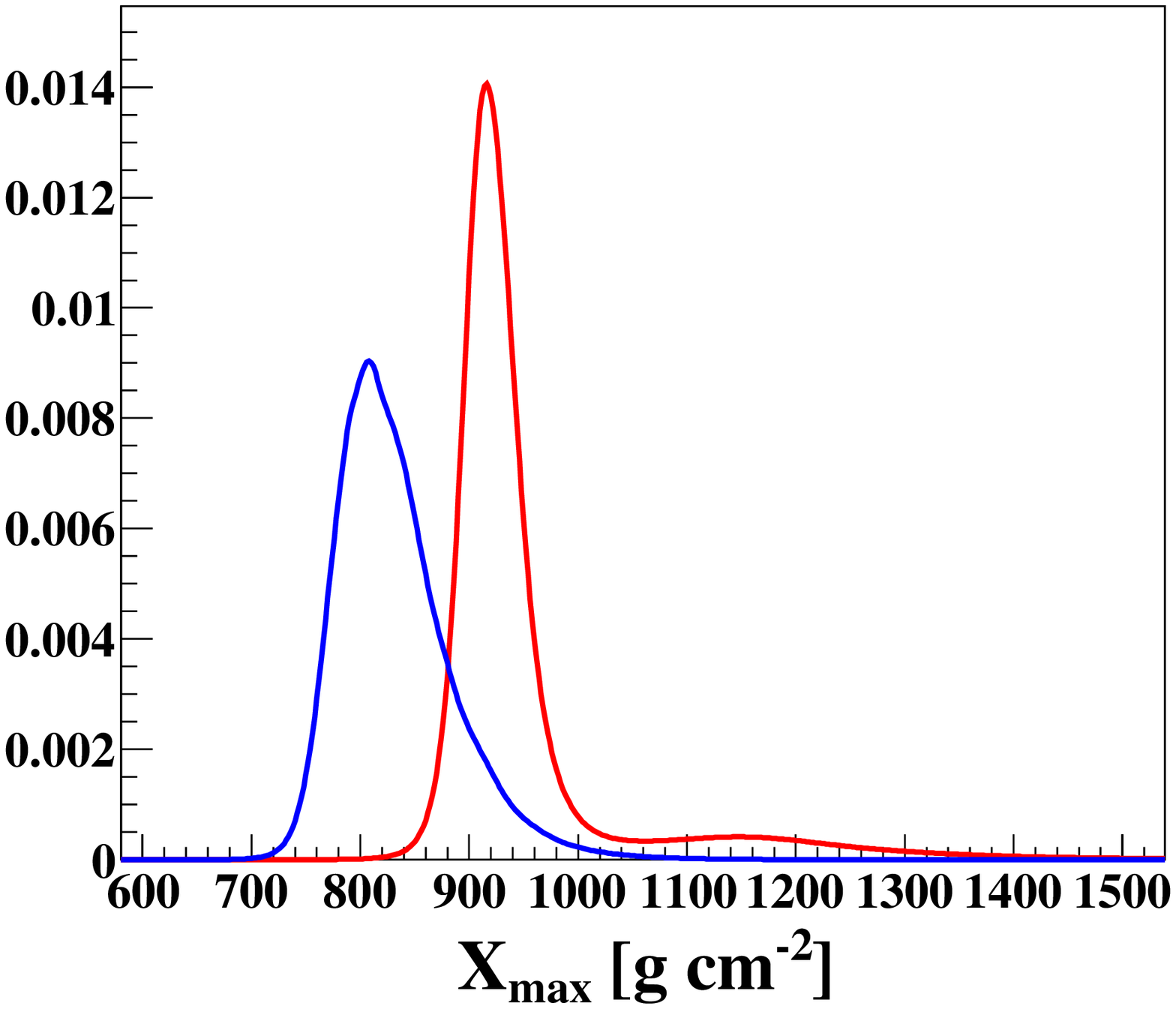}
\caption{Estimates of the protons and photons distribution functions of $X_{max}$ for energy bins of 
$\Delta \log(E/\textrm{eV})=0.1$ wide and for the bin center energy going from $10^{19.35}$ eV (top-left panel) 
up to $10^{20.15}$ eV (bottom-right panel). The showers considered are such that 
$\theta \in [30^\circ, 60^\circ]$.}
\label{PDFEth}
\end{figure}

In order to study the upper limit for a given threshold energy, the distribution functions of $X_{max}$ for protons and 
photons of a given primary energy are obtained from MC data, by using the non-parametric method of kernel superposition 
mentioned above. Fig. \ref{PDFEth} shows the estimates of the proton and photon distribution functions for 
$\theta \in [30^\circ, 60^\circ]$ in energy bins of $\Delta \log(E/\textrm{eV})=0.1$ where the energy corresponding to 
the center of the bin goes from $10^{19.35}$ eV (top-left panel) up to $10^{20.15}$ eV (bottom-right panel). It can be  
seen that, as the energy increases the bump due to the photons that do not convert in the geomagnetic field becomes 
progressively less important. 

The broken power law energy spectrum from Ref. \cite{Inoue:09} is considered in order to obtain the distribution 
functions of $X_{max}$ for protons and photons for a given threshold energy ($E \geq E_{th}$). The number of 
events above a given energy are taken from Ref. \cite{Inoue:09} which corresponds to an orbital detector like 
JEM-EUSO with two years in nadir mode plus three years in tilt mode ($\alpha_{Tilt}=38^\circ$).   

\begin{figure}[!bt]
\centering
\includegraphics[width=11cm]{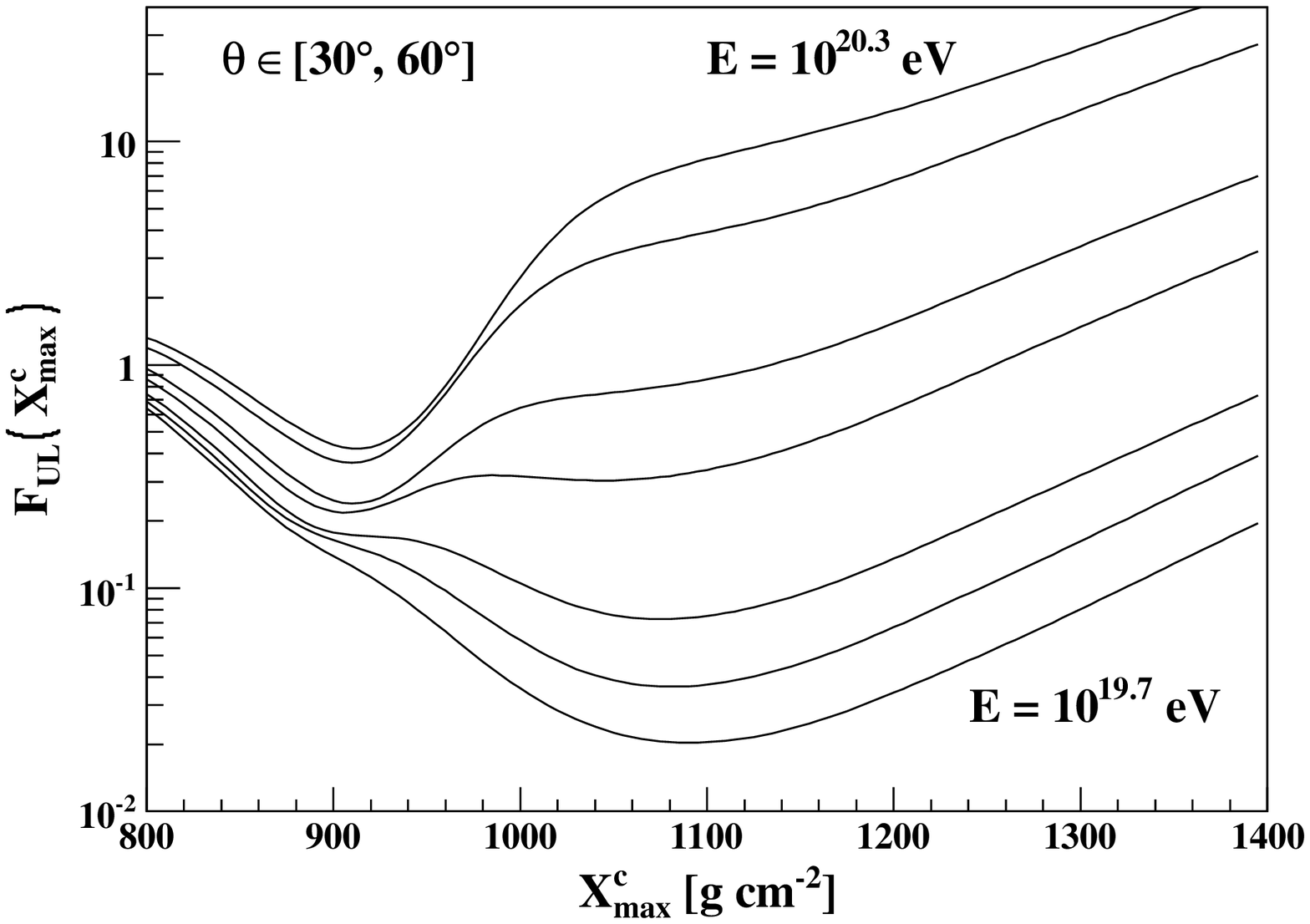}
\caption{Upper limit as a function of $X_{max}^c$ for threshold energies from $10^{19.7}$ eV to $10^{20.3}$ eV 
in steps of $\log(E/\textrm{eV})=0.1$ and for $\theta \in [30^\circ, 60^\circ]$ for the ideal case where 
reconstruction error of $X_{max}$ is zero.}
\label{ULcut}
\end{figure}

Fig. \ref{ULcut} shows $\mathcal{F}_{UL}(X_{max}^c)$ as a function of $X_{max}^c$ obtained by using the distribution 
functions of Fig. \ref{PDFEth} convoluted with the energy spectrum for $\alpha=0.95$ and for 
$\theta \in [30^\circ, 60^\circ]$. Here a 50\% of reconstruction efficiency due to the effects of clouds is assumed
\cite{Berat:10}. It can be seen that $\mathcal{F}_{UL}(X_{max}^c)$ reaches a minimum which depends 
on the threshold energy. Note that there is a transition at $E_{th}\cong10^{20}$ eV from which the minimum is reached 
at larger values of $X_{max}^c$. This is due to the change in the shape of the photon distribution function when the 
threshold energy increases. Finally, the upper limit is obtained evaluating $\mathcal{F}_{UL}$ in $X_{max}^c$ of the 
minimum.    

In order to exemplify the application of these techniques we consider an orbital detector with the
following characteristics:

\begin{itemize}

\item As in the second technique describe above, the energy spectrum and the number of events are obtained 
from Ref. \cite{Inoue:09}. The number of events corresponds to two years in nadir mode plus three years in 
tilt ($\alpha_{Tilt}=38^{\circ}$) mode for the JEM-EUSO mission. 

\item A reconstruction efficiency, taking into account the presence of clouds, of $\epsilon_{R}=50\%$ is 
assumed (see Ref. \cite{Berat:10}). 

\item Although there are estimations of the reconstruction uncertainty of $X_{max}$ at $10^{20}$ eV 
\cite{Berat:10}, it is left as a parameter and the upper limits are calculated as a function of it .  

\end{itemize}

In Ref. \cite{Berat:10} it is shown that the uncertainty on the reconstruction of $X_{max}$ at $10^{20}$ eV
increases with zenith angle going from $\sim 50$ g cm$^{-2}$ at $\theta = 10^\circ$ up to $\sim 90$ g cm$^{-2}$
at $\theta = 80^\circ$. This is obtained without using the information given by the Cherenkov peak, which is 
present in a fraction of events. When the Cherenkov peak is available much better results are obtained, the 
uncertainty in this case goes from $\sim 15$ g cm$^{-2}$ at $\theta = 10^\circ$ up to $\sim 40$ g cm$^{-2}$ 
at $\theta = 60^\circ$. Note that the last method can be used just up to $\theta = 60^\circ$ (see Ref. 
\cite{Berat:10} for details). Inspired in these results and considering that the uncertainty in $X_{max}$ 
increases at lower energies, a Gaussian distribution with $\sigma[X_{max}] \in \{0, 70, 120, 150\}$ g cm$^{-2}$ 
is used in the upper limit calculations.

\begin{figure}[!bt]
\centering
\includegraphics[width=12cm]{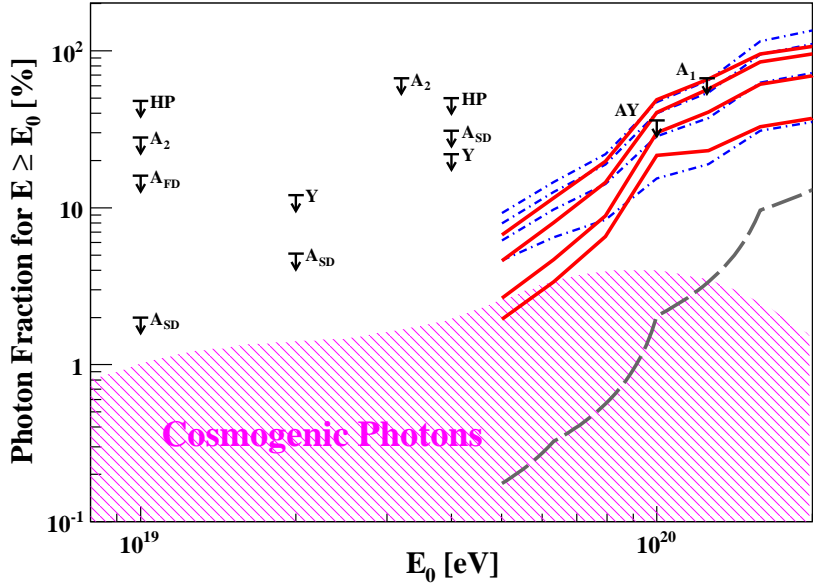}
\caption{The upper limits on the fraction of photons in the integral cosmic ray flux at 95\% of confidence level.
Dashed line corresponds to the ideal case in which it is known that there is no photon in the data. Solid red 
lines are the upper limits obtained by using the optimized cut method. Dash-dot-dash blue lines are the upper 
limits obtained by using $\xi_{X_{max}}$ method. For each method, the lines from bottom to top correspond to a 
Gaussian uncertainty of 0, 70, 120 and 150 g cm$^{-2}$. Shadow region is the prediction for the GZK photons 
\cite{Gelmini:07}. Black arrows are experimental limits, HP: Haverah Park \cite{Ave}; A$_1$, A$_2$: AGASA 
\cite{Risse:05,Shinozaki:02}; A$_{\textrm{FD}}$, A$_{\textrm{SD}}$: Auger \cite{augerFD,augerSD}; AY: AGASA-Yakutsk 
\cite{Rubtsov:06}; Y: Yakutsk \cite{Glushkov:07}.}
\label{ULFinal}
\end{figure}

Fig. \ref{ULFinal} shows the upper limits on the fraction of photons in the integral cosmic ray flux, at 95\%
of confidence level, obtained in the ideal case $\mathcal{F}_{\gamma}^{min}$ (dashed line), by using the optimized 
cut method (solid red lines), by using the $\xi_{X_{max}}$ method (dash-dot-dash blue lines), and also the upper 
limits obtained by different experiments. The calculation is done for $E\geq 5\times10^{19}$ eV and 
$\theta \in [30^\circ, 80^\circ]$. For each method, the lines from bottom to top correspond to a Gaussian 
uncertainty on the determination of $X_{max}$ of 0, 70, 120 and 150 g cm$^{-2}$. 

It can be seen that the upper limits corresponding to $\sigma[X_{max}]=0$, obtained by using both methods 
are about one order of magnitude larger than the ideal case (dashed line in the figure). This is due to 
the limitation imposed by the $X_{max}$ parameter to discriminate between photons and protons. Nevertheless, 
the present analysis can be improved in order to obtain smaller upper limits, much closer to the ideal case. 
This is a work in progress which will be presented in a forthcoming paper.

As expected, the upper limits increases with the uncertainty on $X_{max}$, therefore, it is important to 
improve the existing reconstruction methods and also develop new ones in order to decrease such quantity 
as much as possible. Note that the number of events above a given energy is the only parameter of the 
JEM-EUSO mission used in the present calculation, i.e. given the uncertainty on $X_{max}$ as a function 
of the energy for an orbital observatory with a statistics of the same order of the one of JEM-EUSO, it 
is possible to estimate the attainable upper limits, for the two methods developed here, from the curves 
of Fig. \ref{ULFinal}.   
   
It can also be seen from Fig. \ref{ULFinal} that, in general, the optimized cut method results better than 
$\xi_{X_{max}}$ method, specially for energies bellow $10^{20}$ eV. This happens because the former takes 
advantage of the part of the distribution function originated by photons that do not suffer photon 
splitting. 

It should be noted that, after the comparison with HiRes and Auger data, the energy assigned
to A$_1$ and AY upper limits might be overestimated by as much as 20\%, increasing the
relevance of our technique at energies above $10^{20}$ eV.

\section{Mapping of proton-gamma separation onto the celestial sphere}

\begin{figure}[!bt]
\centering
\includegraphics[width=12cm]{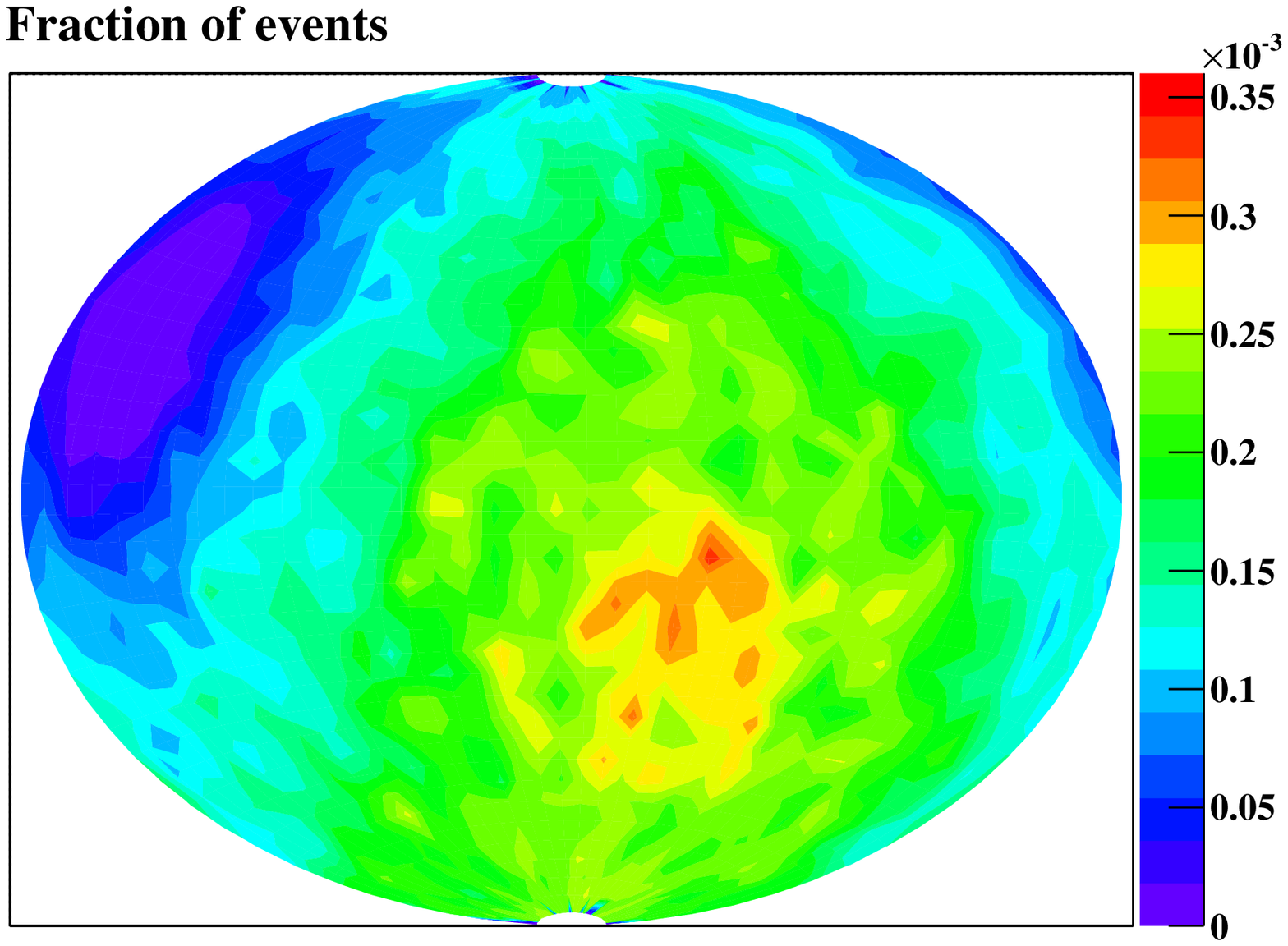}
\includegraphics[width=12cm]{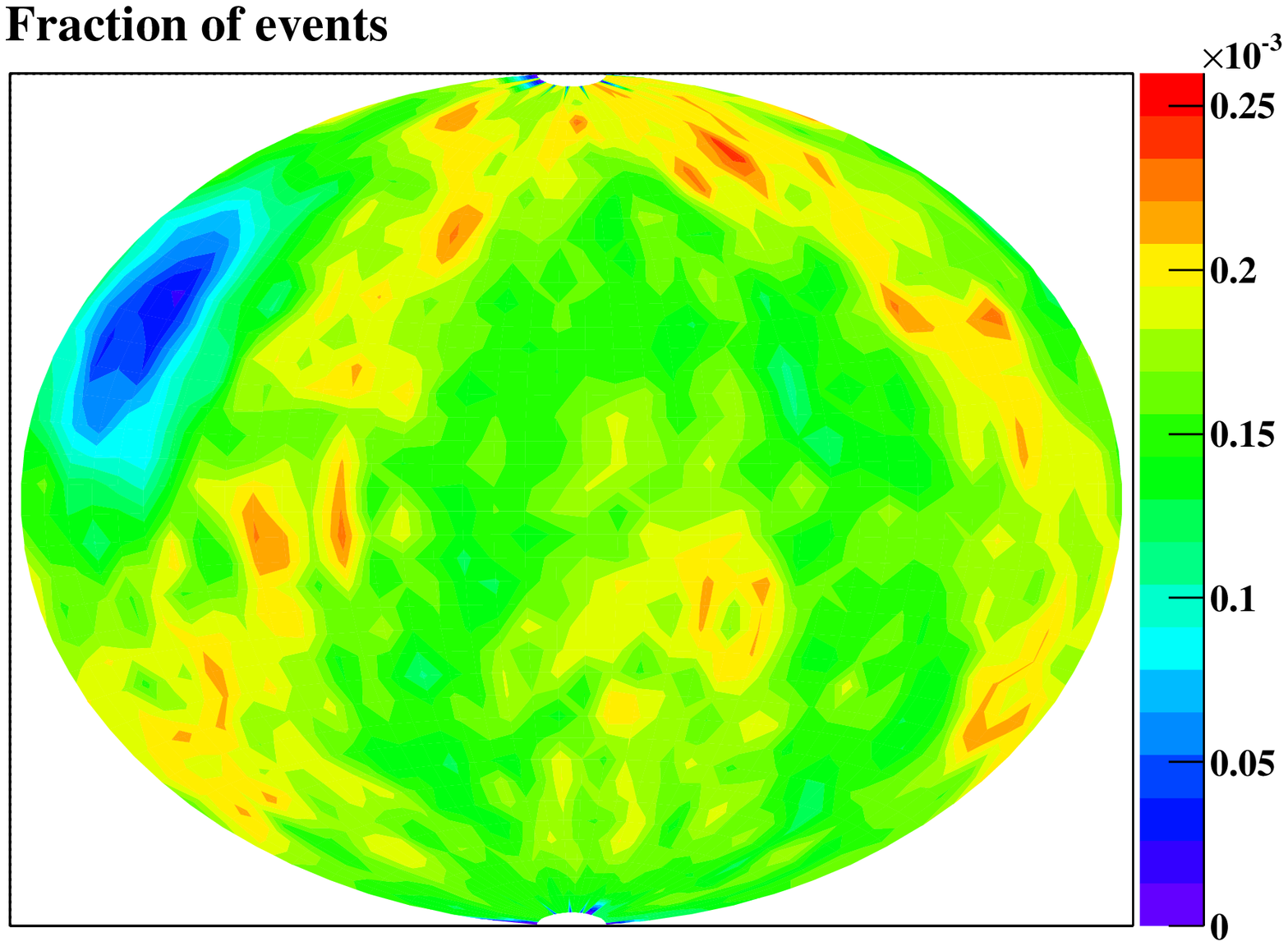}
\caption{Contour plots in galactic coordinates of the fraction of events with $\eta \geq 1.3$ for 
$E \in [10^{19.8}, 10^{20}]$ eV, for $\theta \in [30^\circ, 60^\circ]$ (top panel) and $\theta \in [45^\circ, 90^\circ]$ 
(bottom panel). A circular window of 5$^\circ$ is used for the calculation.}
\label{MapsGal}
\end{figure}

The events that satisfy a given cut in zenith angle and belong to a given mask (which actually depends on such cut) 
come from different regions of the universe. In particular, there are directions in the sky for which the discrimination 
between protons and photons is larger. Fig. \ref{MapsGal} shows maps in galactic coordinates of the fraction of events 
in a circular window of 5$^\circ$ corresponding to two different cuts in zenith angle $\theta \in [30^\circ, 60^\circ]$ 
(top panel) and $\theta \in [45^\circ, 90^\circ]$ (bottom panel) for a mask of $\eta_{Lim}=1.3$. A uniform exposure in 
right ascension is assumed for the calculation.    

From Fig. \ref{MapsGal} it can be seen that for $\theta \in [30^\circ, 60^\circ]$ the fraction of events is larger
than for $\theta \in [45^\circ, 90^\circ]$, which is consistent with the fact that for larger values of zenith
angle the probability of photon splitting is larger. Also the regions on the sky where the discrimination power is
larger corresponds to regions on the Earth surface predominantly in the south hemisphere where the maximum separation
between protons and photons is obtained.

\section{Conclusions}

In this work we study the discrimination between photon and proton showers in the context of future orbital detectors. 
We introduce two complementary techniques that are based on the $X_{max}$ parameter. We find that the discrimination 
power of $X_{max}$ strongly depends on primary energy due to the dependence of the characteristics of the photon 
cascades on the pre-showering effect which becomes more important as the primary energy increases. We also introduce 
the concept of mask which consist in regions over the Earth surface with a given proton-gamma discrimination power. 
In particular, we find that different masks are mapped into different regions of the sky, which means that there are 
specific directions for which the discrimination power is better. Given the limited exposure of any mission, the latter 
has implications that should be accounted for appropriately when applying astrophysical test over the sky. 

We also apply these new techniques to the case of an ideal mission with the same number of events as the one expected 
for JEM-EUSO. We calculate the upper limit on the integrated flux, assuming no photons in the sample, as a function
of the uncertainty on the determination of $X_{max}$. As a result we also obtain, quantitatively, the degradation 
of the performance of our techniques with the reconstruction uncertainty of $X_{max}$, showing the importance 
of the reconstruction methods and their impact on the physical analyses.

The attainable upper limits on the photon fraction obtained by using $X_{max}$ with $\sigma[X_{max}]=0$ is about one 
order of magnitude above the ideal case, in which it is known that there is no photon in the sample. This is due to 
the limitation of the $X_{max}$ parameter to separate proton from gamma showers. New techniques are under study, 
taking advantage of the richness of the spatial observations, in order to obtain such attainable upper limits 
closer to the ideal case.

\section{Acknowledgments}

This work is part of the ongoing effort for the design and development of the JEM-EUSO mission and the definition of 
its scientific objectives. The authors acknowledge the support of UNAM through PAPIIT grant IN115707 and CONACyT through 
its research grants and SNI programs. ADS is supported by a postdoctoral grant from the UNAM.

\end{document}